\documentclass[%
 reprint,
 amsmath,amssymb,
 aps,
 prx,
]{revtex4-2}

\usepackage{graphicx}
\usepackage{dcolumn}
\usepackage{bm}

\usepackage{siunitx}
\DeclareSIUnit\molar{\mole\per\cubic\deci\metre} 

\usepackage{xcolor}
\newcommand{\tsub}[1]{_{\textrm{#1}}}

\newcommand{\CV}[1]{\textcolor{black}{#1}}
\newcommand{\MRG}[1]{\textcolor{black}{#1}}
\newcommand{\MBG}[1]{\textcolor{black}{#1}}
\newcommand{\MBGb}[1]{\textcolor{black}{#1}}
\newcommand{\MBGc}[1]{\textcolor{black}{#1}}
\newcommand{\MBGd}[1]{\textcolor{black}{#1}}
\newcommand{\MRGb}[1]{\textcolor{black}{#1}}
\newcommand{\MRGc}[1]{\textcolor{black}{#1}}

\begin{document}

\preprint{APS/123-QED}

\title{Discovering dynamic laws from observations: \\ the case of self-propelled, interacting colloids}
\author{Miguel Ruiz-Garcia$^{1,2,3,+}$}
\author{C. Miguel Barriuso G.$^{1,3,+}$} 
\author{Lachlan C. Alexander$^4$}
\author{Dirk G. A. L. Aarts$^4$}
\author{Luca M. Ghiringhelli$^5$}
\author{Chantal Valeriani$^{1,3}$}
\email{To whom correspondence should be addressed. E-mail:  miguel.ruiz.garcia@ucm.es, cvaleriani@ucm.es.\\\noindent
$^+$ These authors contributed equally to this work.}



\affiliation{$^1$Departamento de Estructura ed la Materia, F\'isica T\'ermica y Electr\'onica, Universidad Complutense de Madrid, 28040 Madrid, Spain}
\affiliation{$^2$Department of Mathematics, Universidad Carlos III de Madrid, Avenida de la Universidad 30, 28911 Legan\'es, Spain}
\affiliation{$^3$Grupo Interdisciplinar Sistemas Complejos, Madrid, Spain}
\affiliation{$^4$Department of Chemistry, Physical and Theoretical Chemistry Laboratory,  University of Oxford, South Parks Road, Oxford OX1 3QZ, United Kingdom}
\affiliation{$^5$Physics Department and IRIS Adlershof, Humboldt-Universit\"at zu Berlin, Zum Gro{\ss}en Windkanal 6, 12489 Berlin, Germany}%


\date{\today}

\begin{abstract}
Active matter spans a wide range of time and length scales, from groups of cells and synthetic self-propelled colloids to schools of fish and flocks of birds. The theoretical framework describing these systems has shown tremendous success in finding universal phenomenology. 
However, further progress is often burdened by the difficulty of determining  forces controlling the dynamics of individual elements within each system. Accessing this local information is pivotal for the understanding of the physics governing an ensemble of active particles and for the creation of numerical models capable of explaining the observed collective phenomena. In this work, we present ActiveNet, a machine-learning tool consisting of a graph neural network that uses the collective motion of particles to learn  active and two-body forces controlling their individual dynamics. We verify our approach using numerical simulations of active Brownian particles, \CV{ active particles undergoing underdamped Langevin dynamics,} \MBGb{and chiral active Brownian particles} considering different interaction potentials and values of activity. \MRG{Interestingly, ActiveNet can equally learn conservative or non-conservative forces} \MBGb{as well as torques}. \MBGc{
Moreover, ActiveNet has proven to be a useful tool to learn the stochastic contribution to the forces, enabling the estimation of the diffusion coefficients.}
Therefore, all coefficients of the equation of motion of Active Brownian Particles are captured. Finally, we apply ActiveNet to experiments of electrophoretic Janus particles, extracting the active and two-body forces  controlling  colloids' dynamics. 
On the one side, we have learned that the active force depends on the electric field and area fraction. On the other side, we have also discovered a dependence of the two-body interaction with the electric field that leads us to propose that the dominant force between active colloids is a screened electrostatic interaction with a constant length scale. 
We believe that the proposed methodological tool, ActiveNet, might open a new avenue for the study and modeling of experimental suspensions of active particles.
\end{abstract}

\maketitle


\section{Introduction}
\label{sec:intro}

Many living systems are composed of self-propelling (active) elements which interact and generate complex collective phenomena \cite{ramaswamy2010mechanics,vicsek2012collective,marchetti2013hydrodynamics}. Mathematical models of active particles are used to predict the behavior of a plethora of different systems, such as synthetic self-propelled particles \cite{nishiguchi2015mesoscopic,zhang2016directed,yan2016reconfiguring,van2019interrupted,zhang2021active}, groups of living cells \cite{trepat2009physical,volfson2008biomechanical,heinrich2020size,valencia2021interaction,martinez2021active,tah2022minimal,heinrich2021self},  flocks of birds \cite{ling2019behavioural}, schools of fish \cite{becco2006experimental} or even the collective behavior of human crowds \cite{bain2019dynamic}. These inherently out-of-equilibrium systems are characterized by the activity of their individual elements and their inter-particle interactions. Depending on the level of activity and the nature of such interactions, these models are capable of describing phases resembling those in equilibrium  -- solid/crystal, fluid and gas -- or genuinely out-of-equilibrium phases such as living crystalline clusters \cite{palacci2013living,mognetti2013living}, active turbulence \cite{cisneros2010fluid}, motility-induced phase separation (MIPS) \cite{stenhammar2015activity}, self-assembly \cite{murugan2015undesired,mallory2018active} and various types of flocking phases  \cite{narayan2007long,hayakawa2010spatiotemporal,suematsu2010collective}.

\CV{
Modeling systems of active particles has been successful in many cases \cite{toner1998flocks,ramaswamy2010mechanics,vicsek2012collective,marchetti2013hydrodynamics,bi2016motility,alert2020physical}. Several numerical models of active particles have led to the discovery of novel out-of-equilibrium physical behaviours, such as the appearance of a motility-induced phase separated (MIPS) phase in a dense suspension of active Brownian repulsive particles (characterised by a large activity) \cite{stenhammar2015activity}; the appearance of a cluster phase in a dilute suspension of active Brownian attractive particles (characterised by an intermediate activity) \cite{mognetti2013living}; or a transition from a disordered to a flocking phase in a suspension of active aligning particles 
\cite{vicsek2012collective}.  However,  one of the main burdens to the advancement of the field of active matter  has been the difficulty to uncover, in experiments, the correct expressions for the forces controlling the dynamics at the individual particle level. Considering the limitations in comparing numerical results obtained for a suspension of active particles to experimental results obtained for a suspension of active colloids, it might be useful to develop new tools capable of  extracting  active and inter-particle forces directly from the experimental data. Once these forces are \textit{learned}, they will enable us to unravel the physics governing the system's dynamics.
}

We are probably living the golden age of machine learning. As expected, in the last few years machine learning has also had a big impact in active matter, where not only has drastically improved particles' tracking  \cite{max-3d-tracking, rabault2017performing, helgadottir2019digital, Pineda2023} but also led to a growing interest in coupling machine-learning models to active particles, in a quest to mimic the complex behavior of natural systems \cite{gazzola2016learning,verma2018efficient,alkemade2022comparing}. The reverse path has so far been more elusive. Ideally, one would like to use machine learning to extract/learn forces controlling particles' dynamics, giving rise to  complex phenomena. Some works, inspired by the success in passive thermal systems \cite{cubuk2015identifying,schoenholz2016structural,alkemade2022comparing}, have started to explore this latter direction. On the one hand, a machine-learning model has recently been applied to active systems to learn the probability of rearrangements depending on the local structure \cite{tah2021quantifying}, which provides valuable information on the relationship between structure and dynamics, \CV{but does not aim to recover the forces governing the microscopic dynamics}. On the other hand, a recent approach estimates the effective two-body potential from the pair correlation function \cite{bag2021interaction}. Moreover, very recent works have used machine-learning tools to recover models of pairwise particle–cell interactions in mixtures of synthetic particles and biological cancer cells \cite{faria-2022},
to identify the phase transitions of the Vicsek model \cite{Xue_2023}, \MBGd{to characterise  MIPS \cite{MLmips2023}
and to sort microswimmers \cite{torrik2024machine}.} 
Further works have tried to recover the differential equations describing different physical phenomena \cite{brunton2016discovering,rudy2017data,both2021deepmod}, which could be used to build coarse-grained models for active matter.
Without resorting to machine learning, other authors have tackled problems related to ours. Previous works performed averages on stochastic trajectories to estimate the drift and diffusion coefficients of the Fokker-Planck equation \cite{siegert1998analysis,gradivsek2000analysis} or used inverse statistical-mechanical methods to optimize pair potentials reproducing equilibrium many-particle configurations \cite{rechtsman2005optimized,zhang2013probing,wang2020sensitivity}. Other approaches used a basis of {\it a priori} chosen functions to project the dynamics of long stochastic trajectories, extracting force fields and evaluating out-of-equilibrium currents and entropy production in over-damped \cite{frishman2020learning} and under-damped \cite{bruckner2020inferring} systems. 

In our work, we propose a machine-learning approach, ActiveNet, that can be trained to learn the dynamics of a suspension of active particles. The active and inter-particle forces, together with an estimate of the stochastic forces, are directly extracted from our machine-learning tool, once trained using the system's trajectories.
Our method exploits Graph Neural Networks (GNN) \cite{scarselli2009graph,bronstein2017geometric,gilmer2017neural,battaglia2018relational}, which have already been used to study many physical domains \cite{battaglia2016interaction,chang2016compositional,sanchez2018graph,mrowca2018flexible,li2018learning,kipf2018neural,bapst2020unveiling,sanchez2020learning,Xue2022learning}, 
\CV{including physical problems in condensed matter \cite{NatEd2022,Schuetz2022}, material science \cite{Reiser2022,Xie2018}, chemistry \cite{Zeng2022,Wang2022,Chen2022}, soft matter  \cite{Mandal2022,Cheng2022,alkemade2022comparing}, or very recently, even active matter \cite{Pineda2023}. In all applications, the network architecture and the input descriptor have been adapted to the problem under consideration.}  Our approach builds on the concepts and formalism introduced by Cranmer \textit{et al.} \cite{cranmer2019learning,cranmer2020discovering}, who presented the possibility to use GNNs to extract the conservative two-body forces in systems of (a few) passive particles. With the {\it a priori} assumption that, to a first approximation, most systems of active colloids obey an overdamped dynamics, we extend the proof-of-concept work of Refs \cite{cranmer2019learning,cranmer2020discovering} to deal with active forces. 
Differently from \cite{cranmer2019learning,cranmer2020discovering}, 
ActiveNet is capable to tackle systems of thousands of colloids, by means of clustering and sparse graphs.
\CV{In ActiveNet we adopt the {\em ensemble} approach \cite{dietterich2000ensemble} to estimate the error bars of the predicted observables. In practice, a set of GNNs are trained on the same data, each GNN differing due to the 
initial, random guess of its training coefficients. This yields an ensemble of networks that give a distribution of predictions for a given input. The overall prediction of the ensemble is taken as the average of the distribution and the error bar is its standard deviation.}

The goal of ActiveNet is extracting the one-body active force and the two-body interacting forces between pairs of colloids from particles' trajectories. Given that ActiveNet allows one to learn forces (not potentials),   these forces can be  conservative or non-conservative. 
Our approach is not restricted to particles undergoing Brownian dynamics.  ActiveNet can be applied to particles undergoing underdamped Langevin dynamics \cite{Lowen2020,glotzer20}, where forces  depend on  particles' velocities. 
We also show how ActiveNet correctly learns torques, characteristic of chiral active particles \cite{zhan_2022_chiral, Liebchen_2022_chiral}, 
and it is also useful to  estimate stochastic forces. 

All this paves 
the way for learning more complex dynamics present in experiments of colloidal systems where temperature and hydrodynamic interactions might be relevant---the latter usually displaying velocity-dependent forces.

After validating our approach with computer simulations, we use ActiveNet to extract both active and two-body forces from experiments of electrophoretic Janus colloidal particles \cite{squires2006breaking,bukosky2019extreme}. 
Electrophoretic Janus colloids are spherical silica particles  half-coated with titanium, whose self-propulsion is induced by an external electric field creating an asymmetrical dipole on the particles. This  couples to the electrolite where they are submerged, generating an asymmetric ion distribution in the vicinity of each colloid's surface, leading to self-propulsion and complex interactions~\cite{squires2006breaking}. 
With our approach, we are able to independently extract the active and two-body forces  and unravelling the physics dominating their movement. We hope that our approach will lead to new insights in the physics governing experimental active colloids. 


\section{The Graph Neural Network}
\label{sec:gnn}

We start focusing on systems in the overdamped regime (Brownian dynamics), where inertia is neglected and viscous forces dominate the dynamics. 
For this reason, in the following, we will alternatively refer to forces or velocities, since they are related by the Stoke's law (see appendix section \ref{sec:methods_gnn} for more details).
In this limit, the deterministic equation of motion for particle $i$ simplifies to $\dot{\vec{x}}_i \propto \vec{F}_i$, where $\vec{F}_i$ represents the combination of all  forces acting on particle $i$, which can be of different nature, such as inter-particle interactions, external conservative fields or active forces. Active forces model different mechanisms of self-propulsion, where particles (such as bacteria or Janus particles) extract energy from the environment and use it for self-propelling. Although in few simple cases some of the forces could be estimated with other methods --  in diluted cases, the active force can be inferred directly from the average velocity --, the goal of ActiveNet is to disentangle each of the forces directly from the particles' trajectories in situations where several forces are acting together. Once they are learned, these forces can be later on used to predict the dynamics of a new set of particles or to understand the physics governing the dynamics of the system.

Learning to predict the dynamics of active particles can be achieved using a broad range of machine-learning models. For example, one could use a Deep Neural Network (DNN) that takes the positions and internal orientations of all particles as inputs ($3 N$ degrees of freedom for a $2$D system) and predicts velocities ($2N$ degrees of freedom for the same 2D system). Training this model would be very challenging, and even if successful, would lead to a high-dimensional nonlinear function $\vec{f}(\vec{c}_1, \dots, \vec{c}_N): \mathbb{R}^{3N} \to \mathbb{R}^{2N}$, where $\vec{c}_i$ is the coordinate vector of particle $i$ (position and orientation). Even though this network could predict the system's dynamics, it presents two main drawbacks: (i) it cannot be easily applied to a system with a different (varying) number of particles, (ii) there is no guarantee (probably not possible in most cases) that we can use this method to disentangle and extract all forces controlling the dynamics of individual particles, \textit{e.g.} the active and inter-particle forces.

Graph Neural Networks, on the other hand, are combinations of deep neural networks (DNNs) applied sequentially on a graph. In our case, the nodes of the graph are the particles in our experiment or simulation and the edges are the interactions that we  want to study (learn). In particular, two-body interactions will be represented as edges between two nodes in the graph. The DNN that acts on the nodes is referred as node function, whereas the DNN that acts on the edges is called edge function. The edge function uses the information of pairs of particles (e.g., their mutual distance), and leads to the estimate of the two-body force. Whereas the node function takes as input the coordinates of a particle and the output of the edge function (see appendix section \ref{sec:methods_gnn} for more details) and will lead to the estimate of the one-body forces, such as the active force, \MBGb{and if it is the case, the drag force or the torque}. 
In practice, for each time frame, a new graph is created (if it is the case, with a different number of nodes), and the GNN (defined by the functions applied to the nodes and edges) is applied. The modularity of the GNN and the possibility of adding inductive biases -- \textit{a priori} assumptions  simplifying the model -- make them easier to train.  Even more, by design \CV{the same GNN can be applied}
to systems of different number of particles, and after training one can easily extract all forces acting on each particle.

Compared to the naive approach of adopting a single, huge neural network, trained over all data, the GNN approach is less complex. The lower complexity is made possible by some crucial, physically justified assumptions:  a) the same forces control the dynamics of all particles (particles are identical)\MBGb{---only one node and one edge function are applied to the graph, b) the total force acting on a particle is the sum of the individual forces acting on it, and c) different modules in the GNN learn different forces, in our case we use the node function for the one-body forces (active forces, drag forces or torques) and the edge function for the two-body forces; 3-body, \dots, N-body interactions can be introduced naturally using new modules in the GNN.} 

\CV{The training of the GNN goes as follows. First we apply the edge function $\vec{\xi}(\vec{c}_i,\vec{c}_j)$ to a node $i$ and each of its $j$-neighbors (each pair of particles constitutes an edge, $ij$). Then, we sum up these outputs and  feed the result to the node function, $\vec{\psi}$, along with the coordinates of the $i$-th node, $\vec{c}_i$. The output of the node function is the predicted velocity for the $i$-th particle $\vec{v}_i^{p}$. Mathematically,}
\begin{equation}
    \vec{v}_i^{\,p} \equiv \vec{\psi} \left(\vec{c}_i,\sum\limits_{d_{ij}<\Gamma} \vec{\xi}(\vec{c}_i,\vec{c}_j) \right)
\end{equation}
where $d_{ij}$ is the Euclidean distance between particles $i$ and $j$. Since we tackle systems of thousands of particles and the number of edges in a fully connected graph scales as $\sim N^2$, we introduce edges in our graph only between pairs of particles such that $d_{ij}<\Gamma$: the number of edges scales now as $\sim N$. \CV{This process is repeated for each particle in the system, using the same node and edge functions. The training is performed by} 
minimizing the difference between the predicted and \MBG{ground-truth} velocities using an $L^1$ norm as loss function:
\begin{equation}
    \mathcal{L} = \sum\limits_i |\vec{v_i}-\vec{v_i}^p| 
\end{equation}

In figure \ref{fig:GNN}, we show a diagram of the basic idea behind the GNN workflow, that we will name ``ActiveNet''.
\begin{figure}[h!]
    \centering
  \includegraphics[width=\linewidth]{FIGURE_2.pdf}
    \caption{
    \textbf{Predicting the dynamics of active particles with a Graph Neural Network (ActiveNet) while learning the functional form of the active and inter-particle forces. }
    (a) ActiveNet is formed by a node function $\vec{\psi}$ (in pink) and an edge function $\vec{\xi}$ (in orange). Function $\vec{\xi}$ takes the coordinates $\vec{c}$ of two particles and, after training, it outputs a linear transformation of the two-body force acting between them. Function $\vec{\psi}$ takes the coordinates of particle $i$ and the sum of the outputs of $\vec{\xi}$ for all the edges $ij$ such that $|\vec{r}_i-\vec{r}_{j}|<\Gamma$. After training, the output of $\vec{\psi}$ ($\vec{v}_i^p$) is the predicted velocity of particle $i$ (acceleration in the case of underdamped dynamics). Applying ActiveNet to all particles in the system (the graph) provides all the predicted velocities. During the learning process the internal parameters of $\vec{\xi}$ and  $\vec{\psi}$ are optimized so that all the $\vec{v}_i^p$ approach the \MBG{ground-truth} velocities. \CV{Both the node (b) and edge (c) functions are neural networks with two hidden layers of 300 neurons and} the appropriate input and output dimensions.}
    \label{fig:GNN}
\end{figure}

\CV{Note that this scheme can be also applied to the case of underdamped dynamics, where ActiveNet outputs the predicted acceleration instead of the predicted velocity of the particle. In this case, velocities can be used as inputs to $\vec{\xi}$ and $\vec{\psi}$, making it possible to learn forces that depend on velocities.} Once ActiveNet is trained, we use $\vec{\xi}$ and $\vec{\psi}$ to extract the inter-particle and active forces that it has learned (see the appendix section \ref{sec:methods_gnn} for more details). Note that the two-body force learned for distances larger than $\Gamma$ will be meaningless due to the absence of data. In practice, we choose a small $\Gamma$ value for a first training, and train again the model with larger values of $\Gamma$ until we see that the inter-particle force term goes to zero. Thus, we gain no information by increasing $\Gamma$ even further.


In the following sections, we test ActiveNet against numerical simulations where we can compare  forces that the model learns with the ground truth used in  simulations. Once we validate our approach, we use ActiveNet to study  forces present in an experimental system of active particles.

\section{Numerical simulations details}
\label{sec:methods_md}

\begin{figure}
\centering
\includegraphics[width=0.8\columnwidth]{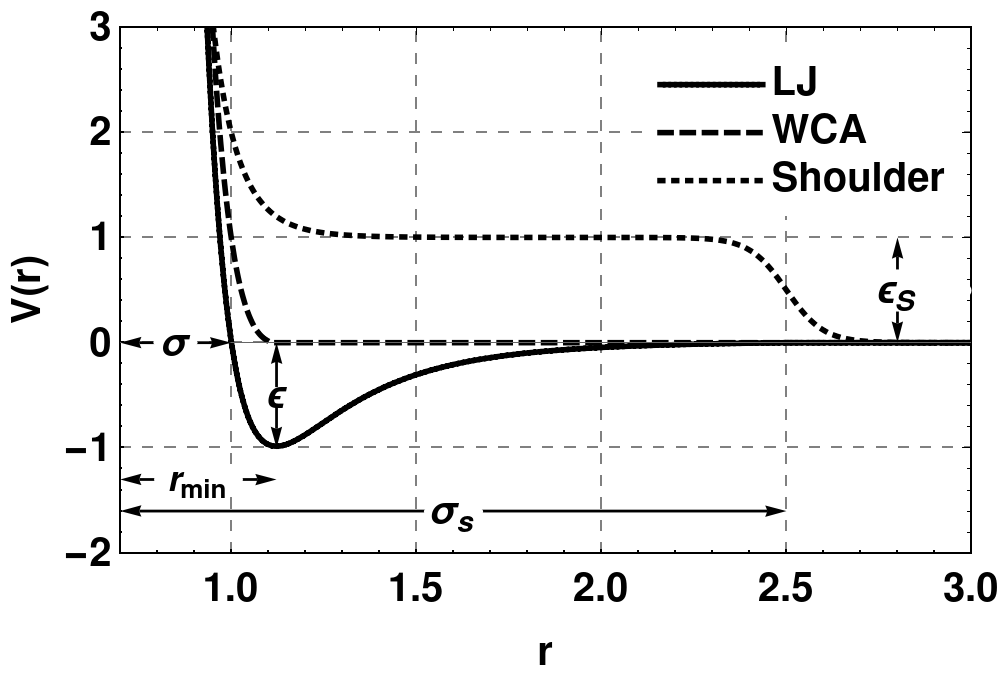}
\caption{{\bf Two-body potentials used in the numerical simulations}. The Lennard-Jones potential describes short-range attraction between particles. 
WCA and shoulder potentials lead to purely repulsive forces between  particles. ActiveNet can learn the inter-particle passive force derived from all three potentials. }
\label{fig:potentials}
\end{figure}

Throughout our study, we consider three interaction potentials  between particles ($V(r_{ij})$, see figure \ref{fig:potentials}): the truncated and shifted Lennard-Jones potential (LJ) and two  repulsive potentials,  WCA \cite{WCA}  and a shoulder potential \cite{gribova2009waterlike}. The short-range attractive Lennard-Jones potential obeys the following equation
\begin{equation}
    V_{\text{LJ}}(r)  = 
        4 \epsilon \left[ \left( \dfrac{\sigma}{r} \right)^{12} -  \left(\dfrac{\sigma}{r} \right)^{6} \right],
     \label{eq:LJ}
\end{equation}
where $r$ is the center-to-center distance, $\sigma$ is the particle diameter and $\epsilon$ is the depth of the minimum, that quantifies the attraction strength. The truncated and shifted LJ potential is accomplished by truncating $V_{\text{LJ}}$ to zero after a given cutoff ($r_{\text{cut}}=2.5\sigma$), and shifting it up to recover continuity at $r=r_{\text{cut}}$:
\begin{equation}
    V_{\text{TSLJ}}(r)  = \left\{ \begin{array}{lcl}
      V_{\text{LJ}}(r) - V_{\text{LJ}}(2.5\sigma) & , & r< 2.5  \sigma \\
        0 & , & r \geq 2.5  \sigma
    \end{array}  \right.
     \label{eq:TSLJ}
\end{equation}
The repulsive WCA potential can be written as a truncated LJ potential  setting $r_{\text{cut}}=r_{\text{min}}=2^{1/6}\sigma$ (the distance to the potential minimum), and shifting it up to recover continuity at $r=r_{\text{cut}}$:
\begin{equation}
    V_{\text{WCA}}(r)  = \left\{ \begin{array}{lcl}
        V_{\text{LJ}}(r) +\epsilon & , & r< 2^{1/6}  \sigma \\
        0 & , & r \geq 2^{1/6}  \sigma
    \end{array}  \right.
     \label{eq:WCA}
\end{equation}
where $V_{\text{LJ}}(r_\text{min})=-\epsilon$. 
Finally, the repulsive  shoulder potential is characterized by two different length scales: a repulsive hard core and a soft repulsive shell around each particle (see figure \ref{fig:potentials}).
According to \cite{gribova2009waterlike},
\begin{equation}
    V(r)=\epsilon \left( \frac{\sigma}{r} \right)^n + \frac{1}{2} \, \varepsilon_s \, \left\{ 1-\tanh{\left[ k_0 \, \left( r- \sigma_s \right) \right]}\right\},
    \label{eq:shoulder_potential}
\end{equation}
where $\sigma$ is the hard core diameter, $\epsilon_s$ and $\sigma_s$ are the height and width of the repulsive shoulder, respectively, $n$ affects the stiffness of the repulsive core and $k_0$ describes the steepness of the shoulder decay (figure \ref{fig:potentials}).  
Following reference [\citenum{gribova2009waterlike}],  we use the following parameters: $n=14$, $k_0 =10/\sigma$  and $\sigma_s=2.5$.


In all simulations we set $\epsilon=\sigma=m=1$ and the quantities are expressed in reduced LJ units, $r^*=r/\sigma$, $\tau^*=\tau\sqrt{\epsilon/m\sigma^2}$, $F^*=\sigma F/\epsilon$ and $U^*=U/\epsilon$ and $k_B=1$. Throughout the manuscript, we drop the asterisks to avoid cluttering all equations.\\

To test ActiveNet in a wide range of case studies we simulated a two-dimensional suspension of active particles undergoing \MBGb{three different dynamics: (i) active Brownian dynamics, (ii) active underdamped Langevin dynamics and (iii) chiral active Brownian dynamics}.\\

\MBGb{(i)} We perform Brownian dynamics simulations, with an in house modified version of the {\it LAMMPS} \cite{LAMMPS} open-source package.
We simulate $N=2\,500$ circular particles with diameter $\sigma$ for $T_\text{sim}=10^7$ timesteps with $\Delta t=10^{-5}$ (reduced units) \cite{Stenhammar}, in a two-dimensional box of size $L \times L$  (with periodic boundary conditions) such that the desired total number density $\rho= \frac{N}{L}\in\{0.1,0.3,0.6\}$ \footnote{As in reference [\citenum{rogel}], we use the total density of the system instead of the packing fraction, since a particle's diameter (necessary for computing the packing fraction) might not be uniquely defined due to particles' activities (and cannot be estimated via the Barker and Henderson's approach \cite{barker}).}. The equations of motion for the position $\vec{r}_i$ and orientation $\theta_i$ of the $i$-th active particle can be written as:
\begin{align}
& \dot{\vec{r}}_i = \frac{D_t}{k_\textrm{B} T} \left( - \sum_{j\neq i} \nabla V(r_{ij}) +  F_p \, \vec{n}_i \right) + \sqrt{2D_t} \, \vec{\xi}_i,
\label{ec:abp_tras}\\
& \dot{\theta}_i = \sqrt{2D_r}\, \xi_{i,\theta},\label{eq:abp_rot}
\end{align}
where $V(r_{ij})$ is the inter-particle pair potential, $k_B$ the Boltzmann constant, $T$ the absolute temperature, $F_p$ a constant self-propulsion force acting along the orientation vector $\vec{n}_i$, which forms an angle $\theta_i$ with the positive $x$-axis, $D_t$ and $D_r$ are the translational and rotational diffusion coefficients. The components of the thermal forces $\vec{\xi}_i$ and $\xi_{i,\theta}$ are white noise with zero mean and correlations $\langle\xi^{\alpha}_i(t)\xi^{\beta}_j(t')\rangle = \delta_{ij} \delta_{\alpha\beta} \delta(t-t')$, where $\alpha,\beta$ are the $x$, $y$ components, and $\langle\xi_{i,\theta}(t)\xi_{j,\theta}(t')\rangle = \delta_{ij} \delta(t-t')$.  \CV{We set $D_t=k_B T/\gamma=0.01$, with $\gamma = 1$}, $D_r\in\{0.25,1.0\}$ and $F_p\in\{3,15,30,60,120\}$ for three different $V(r_{ij})$ (WCA, Lennard-Jones and shoulder-like potential shown in figure \ref{fig:potentials}) in order to achieve a wide range of phases, although we only included three cases in the main text.\\

\MBGb{(ii)} \CV{ Next, we  simulate  active  particles undergoing underdamped Langevin dynamics with the {\it LAMMPS} \cite{LAMMPS} open source package, for $T_\text{sim}=10^7$ timesteps with $dt=10^{-3}$ (reduced units). }
\CV{The equations of motion for the position $\vec{r}_i$ and orientation $\theta_i$ of the $i$-th active particle can be written as:
\begin{align}
\label{eq:motion2}
& m \dot{\vec{v}}_i =  - \sum_{j\neq i} \nabla V(r_{ij}) +  
F_p \, \vec{n}_i  - \gamma_t \vec{v}_i +
\sqrt{2\gamma_t k_B T} \, \vec{\xi}_i, \\
& \dot{\theta}_i = \sqrt{2D_r}\, \xi_{i,\theta}
\end{align}
where $V(r_{ij})$ is the inter-particle pair potential, $F_p$ a constant self-propulsion force acting along the 
orientation vector $\vec{n}_i$ which forms an angle $\theta_i$ with the positive $x$-axis, $\gamma_t$ the translational friction coefficient, 
and $D_r = k_BT/\gamma_r$ the rotational diffusion coefficient. Furthermore, the components of the thermal forces $\vec{\xi}_i$ and $\xi_{i,\theta}$ are white noise with zero mean and correlations $\langle\xi^{\alpha}_i(t)\xi^{\beta}_j(t')\rangle = \delta_{ij} \delta_{\alpha\beta} \delta(t-t')$, where $\alpha,\beta$ are the $x$, $y$ components, and $\langle\xi_{i,\theta}(t)\xi_{j,\theta}(t')\rangle = \delta_{ij} \delta(t-t')$. }
\CV{We simulated this underdamped system  with 3\,600 active particles with a repulsive WCA potential (equation \ref{eq:WCA}) and $F_p=5$ in a rectangular box with an edge of $L=600$ corresponding to a density of $\rho = 0.01$. The temperature was set to $k_B T=1$ with a traslational and rotational friction coefficients of $\gamma_t = 1$ and $\gamma_r = 10/3$ respectively, consistent with the rotational diffusivity of spherical particles \cite{glotzer20}.}\\

\MBGb{(iii) Finally, we perform chiral active Brownian simulations, similarly to (i) with an in house modified version of the LAMMPS \cite{LAMMPS} open-source package. In this case the equations governing this system are the same as (i) equations (\ref{ec:abp_tras}) and (\ref{eq:abp_rot}) but we add a constant torque, which in the overdamped regime leads to a constant term in equation (\ref{eq:abp_rot}), a constant angular velocity $\omega_0$ for the orientation of the particle: $\dot{\theta}_i = \omega_0 + \sqrt{2D_r}\xi_{i,\theta}$. Using the angular version of Stokes' law,  $\mathcal{T} = 8\pi\eta R^3\omega$ and the rotational Stokes-Einstein relation $D_r = k_B T / 8\pi\eta R^3$,  we can write $\mathcal{T} = \left(k_B T/D_r\right)\omega$. So equation (\ref{eq:abp_rot}) is rewritten as,
\begin{equation}
    \label{eq:cabp_rot}
    \dot{\theta}_i = \frac{D_r}{k_B T}\mathcal{T}_0 + \sqrt{2D_r}\xi_{i,\theta}
\end{equation}
now showing the constant torque $\mathcal{T}_0$. In the case of no interaction and null noise, the orientation of these particles rotates with a constant $\omega_0$ and the particles perform circular motions with radius $R_g = v_0/\omega_0$ and period $T_g = 2\pi/\omega_0$. We perform simulations of 2\,500 cABPs described by equations (\ref{ec:abp_tras}) and (\ref{eq:cabp_rot})  for $T_\text{sim} = 10^7$ timesteps with $dt=10^{-5}$, in a square simulation box of side $L\approx 91.29$ setting the numerical density at $\rho = 0.3$. For the inter-particle interactions we use a WCA potential (\ref{eq:WCA}). We set $k_B T = D_t = 10^{-4}$, $D_r = 6.28\cdot 10^{-5}$ and $F_p = 3.14\cdot 10^2$ to perform 5 simulations for $\mathcal{T}_0 = \{10^{-1}, 10^{0}, 10^{1}, 10^{2}, 10^{3}\}$ leading to  $R_g = 5\sigma\cdot\{10^{-1}, 10^{0}, 10^{1}, 10^{2}, 10^{3}\}$ and  $T_g \approx \{10^{-2}, 10^{-1}, 10^{0}, 10^{1}, 10^{2}\}/dt$ steps.}\\


In the case of Brownian dynamics, to train our neural network, we use the position and orientation of all particles at different times as input data and the velocities as the variables that the GNN should reproduce at its output. At time $t$ we calculate the ground-truth velocities as $\left[\vec{x}(t+\Delta t) - \vec{x}(t)\right]/\Delta t$, setting $\Delta t = 10$ simulation steps. We have checked that the GNN obtains equivalent results choosing $\Delta t = 10, 50, 100$. In general, using a smaller $\Delta t$ will lead to a better correlation with the instantaneous forces present at $t$ but noisier data (more data may be necessary to train). On the other hand, a larger $\Delta t$ will lead to a larger signal-to-noise ratio (less data may be necessary), but a degraded correlation with the force we want to learn. 
\CV{In the case of underdamped Langevin dynamics, to train our neural network, we use position, velocity and orientation of all  particles at different times as input data and the accelerations as the variables that the GNN should reproduce at its output. At a time $t$ we calculate the accelerations as $\left[\vec{v}(t+\Delta t) - \vec{v}(t)\right]/\Delta t$, setting $\Delta t = 10$ simulation steps. } 
Finally, in the case of chiral active particles, we  calculate the angular velocity as $\left[\theta_s(t+\Delta t)-\theta_s(t)\right]/\Delta t$, where $\theta_s$ is the total swept angle by the orientation of the particle with respect to the positive $x$-axis at $t=0$ and $\Delta t=100$ simulation steps.

\section{Experimental details}

\label{sec:methods_exp}

\begin{figure}
    \centering
    \includegraphics[width=0.75\linewidth]{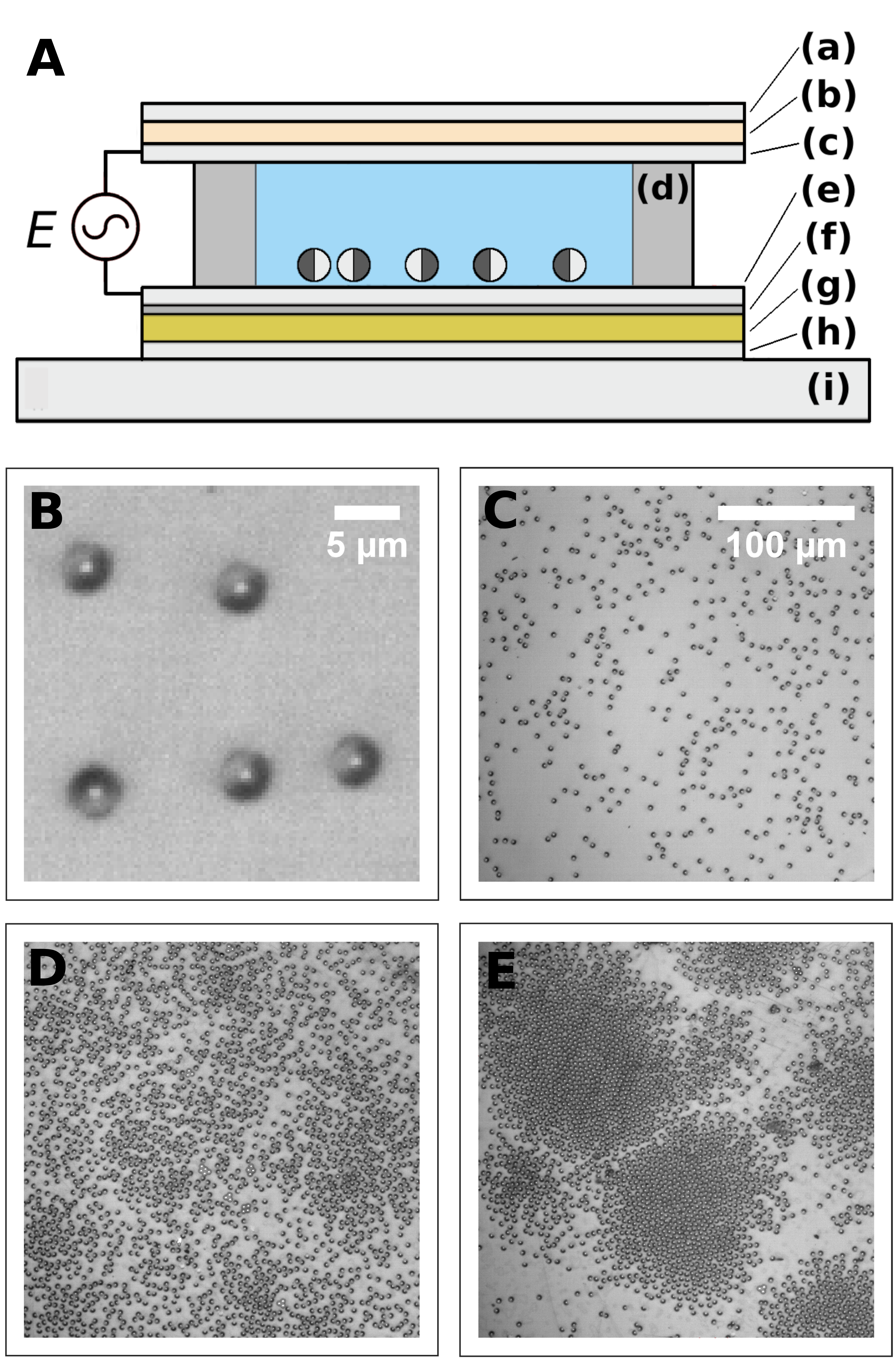}
     \caption{
   {\bf Experimental set-up}. (A) Scheme of an observation cell: (a) Cover slip, (b) ITO electrode, (c) 25 nm silica surface, (d) spacer with UV glue, (e) 25 nm silica surface, (f) 5 nm chromium surface, (g) 80 nm gold surface, (h) cover slip and (i) glass slide. \CV{(B-E) Snapshots of the experimental system. (B) Zoom on five electrophoretic Janus particles where the two hemispherical caps are clearly seen. When an electric field, $E$, is applied perpendicular to the substrate, particles orient themselves in such a way to maximize the magnitude of their induced dipoles. The equator that separates the caps is perpendicular to the substrate, confining the movement to $2$D. Depending on the intensity of the electric field, $E$, and the area fraction, $\phi$, different phases can be observed: (C) a gas phase (dilute, $\phi = 0.10$, $E = 727$ Vcm$^{-1}$); (D) a liquid-like phase  ($\phi = 0.29$, $E = 181$ Vcm$^{-1}$, where activity is too low for phase separation to take place);  (E) an interrupted phase separation ($\phi = 0.29$, $E = 363$ Vcm$^{-1}$).}}
    \label{fig:exp_snapshots}
\end{figure}

We have performed experiments in a quasi-2D system of induced-charge electrophoretic self-propelled Janus colloids \cite{nishiguchi2015mesoscopic,yan2016reconfiguring,zhang2016directed}. Due to the AC field, colloids self-propel and exhibit inter-particle interactions also, as a result of their electric polarization.

We used silica particles with a diameter of \SI{4.28}{\micro \metre} to create Janus particles. We first deposited the silica particles onto cleaned glass slides with a resulting area fraction of \num{0.1}, after which we left the solvent evaporate. The particles were then coated with \SI{50}{\nano \metre} of titanium by vapour deposition, and subsequently coated with \SI{15}{\nano \metre} of silica. These particles were removed from the cover slip by gentle sonication into a NaCl \SI{0.1}{\milli \molar} solution. 

The sample cell was built with the following specifications: top electrode \num{15}-\SI{30}{\ohm \per \centi \metre} ITO cover slips from Diamond Coatings Ltd, coated with \SI{25}{\nano \metre} silica by vapour deposition;  bottom electrode: \SI{80}{\nano \metre} gold electrode with \SI{5}{\nano \metre} of chromium and \SI{25}{\nano \metre} of silica. The bottom part is in contact with particles. Specac Omni Cell spacers from Merck of width \SI{60}{\micro \metre} were used with Norland optical UV glue to separate the electrodes. This left a gap of \SI{110(010)}{\micro \metre} between the electrodes. An alternating square potential at \SI{8}{\kilo \hertz} and varying amplitude was passed through a signal generator creating a field perpendicular to the observation plane. \CV{A sketch of the experimental set-up is shown in figure \ref{fig:exp_snapshots}.}

Recordings were made with a $4.2MPix$ XIMEA camera at a constant framerate ($<10.3fps$) in a reflection microscopy setup. To this end, a BS013 \num{50}:\num{50} beam splitter from Thorlabs was used with an Olympus UPLXAPO 20X oil immersion Objective. Particles were detected with a custom algorithm and tracked with software by Crocker and Grier \cite{Crocker1996}. 

The behavior of electro-phoretic Janus particles is tunable. Nishiguchi et al. \cite{nishiguchi2018flagellar} showed that the direction and speed of the particles could be changed by increasing the frequency of the AC electric field. Zhang et al. and Yan et al. also demonstrated that the electrostatic interactions between the induced charges in the particles change and invert through a similar change in electric field frequency \cite{zhang2016directed,yan2016reconfiguring}. In our work, we selected the electric field and particle properties such that we avoided the regime where forces between the metal and dielectric cap were attractive, aiming for purely repulsive forces through which we could form MIPS. Note that the conductivity of ions in solution will typically be lower than that of electrons in the metallic cap, since the ions will interact with the solvent molecules and other ions. As a result, the velocities of our particles became quenched when the frequency of the electric field entered the MHz region, and visible reduction already occurred in the kHz range. The screening effects of the ions/double layers, which respond to the induced polarization of the particle, may thus be more isotropic than if the ions would move in sync with the AC field.

Regarding the inertial behaviour of the particles it should be noted that most experiments with passive colloids have been modeled in the overdamped (Brownian) regime \cite{vissers2011,vissers2011,cates2009}. For active colloids, hydrodynamics is often considered to be relevant for unraveling particles' motion. Photocatalytic 
TiO$_2$-functionalised Janus microswimmers, self-propelling when exposed to ultra-violet light \cite{Bailey2022}, have shown complex two- and three-dimensional motion, controlled by the hydrodynamic interactions of the colloids with the glass substrate \cite{Uspal2015}. In \cite{Vutukuri2020}, the authors studied  half-gold-coated TiO$_2$ particles, whose direction of motion could be reversed by exploiting the different photocatalytic activities on both sides. The reversal in propulsion direction changed the hydrodynamic interaction from attractive to repulsive,  qualitatively described  by a minimal hydrodynamic model.
However it is also often common to map experiments of active colloids onto simulation results of active Brownian particles (neglecting, to a first approximation, hydrodynamics). Just to give a few examples, experimental results of active colloids have been compared to numerical results of active overdamped Brownian particles in reference~\cite{Carrasco2023} (Janus Platinum-Polysterene catalytic microswimmers with tunable buoyant weight), in~\cite{Ramananarivo2019} (light-activated microswimmers, with an inserted hematite), in~\cite{Narinder2021} (silica spheres half-coated with a carbon layer in a critical fluid), and in \cite{Fernandez2020} and \cite{yan2016reconfiguring} (induced-charge electro-phoretic colloids in an ac field). In the latest work~\cite{yan2016reconfiguring}, the authors employed the same experimental set-up as the one  used in our work.
Thus, to a first approximation, in the dilute regime and for low electric field amplitudes under study, we assume that the dynamics of the Janus colloids used in our experiments can be considered overdamped.


\section{Results}
\label{sec:results}

\MRGb{In this section, we start by showing that ActiveNet can learn  forces used in simulations of active Brownian particles. Interestingly, we demonstrate that ActiveNet can also learn the stochastic forces present in the dynamics. Next, we show that ActiveNet is able to learn torques (for a system of chiral active Brownian particles) and non-conservative forces (in underdamped dynamics). 
Finally, we demonstrate how ActiveNet learns the forces present in an experimental system and discuss the physical implication of their functional dependence on external parameters such as the electric field}.

\subsection{ActiveNet correctly learns forces in simulations of active particles}

\begin{figure*}
\centering\includegraphics[width=\linewidth]{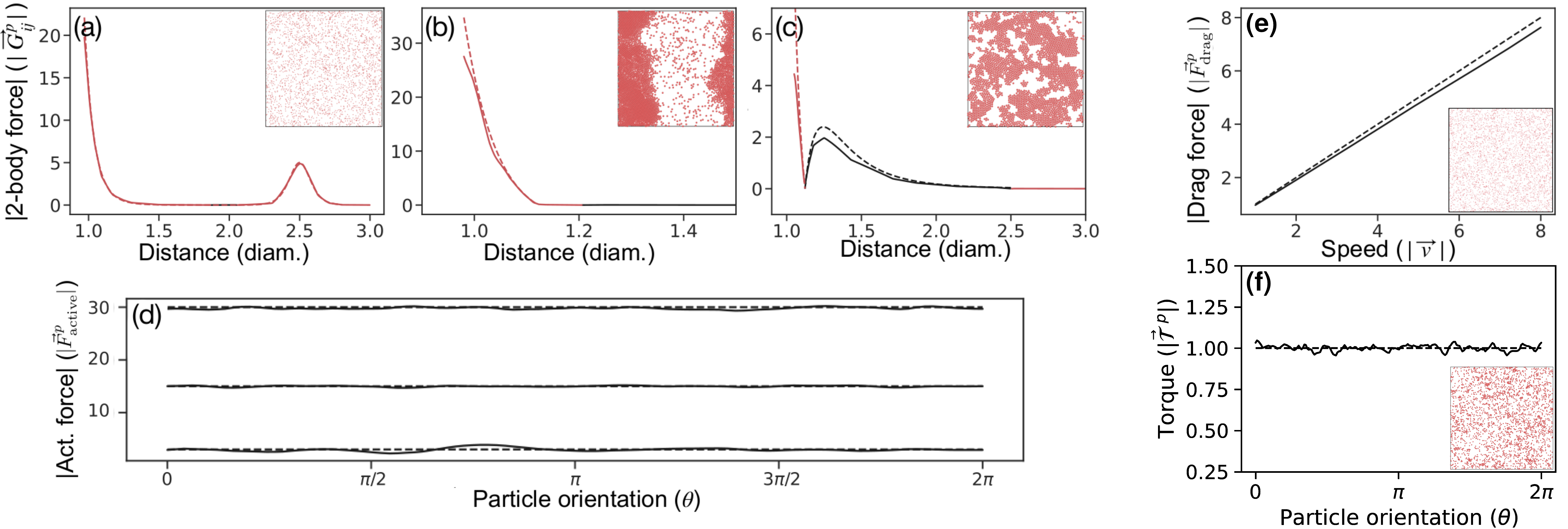}
    \caption{\CV{{\bf ActiveNet can learn the forces \MBGb{and torques} present in  simulations of active particles from their trajectories}. To test ActiveNet, we perform computer simulations on two-dimensional suspensions of different systems. \CV{Panels (a), (b) and (c) illustrate the two-body force versus the distance between two particles in several cases: (a) shows the case of a shoulder-like potential for active particles in a gas phase ($F_p=15$); panel (b) corresponds to active repulsive (WCA-like) Brownian particles in a MIPS phase ($F_p=30$); and panel (c) displays the case of attractive (Lennard-Jones) active Brownian particles in a dynamic cluster phase ($F_p=3$). Panel (d) depicts the magnitude of the active force acting on each particle versus the angle $\theta$ for the three previous cases, either enforced in the simulations (dashed line) or learned from ActiveNet as in \eqref{eq_act_force} (continuous line). The lowest lines correspond to $F_p=3$ (LJ), the middle lines to $F_p=15$ (shoulder) and the top lines to $F_p=30$ (WCA). Panel (e) shows the drag force versus the particle's velocity for active repulsive (WCA-like) particles undergoing underdamped Langevin dynamics in a gas phase. \MBGb{Panel (f) shows the learned torque in the case of chiral ABPs subjected to a constant torque ($\mathcal{T}_0 = 1$ in this case) with a WCA potential, see sections \ref{sec:methods_md} and \ref{sec:methods_gnn} and figure \ref{fig:chiral_learning} for more details}. In all panels the \MBG{ground-truth} force (the one inputted in the simulations) is plotted with dashed lines, whereas the predictions of ActiveNet are plotted with continous lines. In panels (a), (b) and (c) repulsive interactions are plotted in red and attractive in black. The insets show snapshots of the corresponding simulations.}}}
    \label{fig:gnn_MD}
\end{figure*}

\begin{figure}
\centering\includegraphics[width=\linewidth]{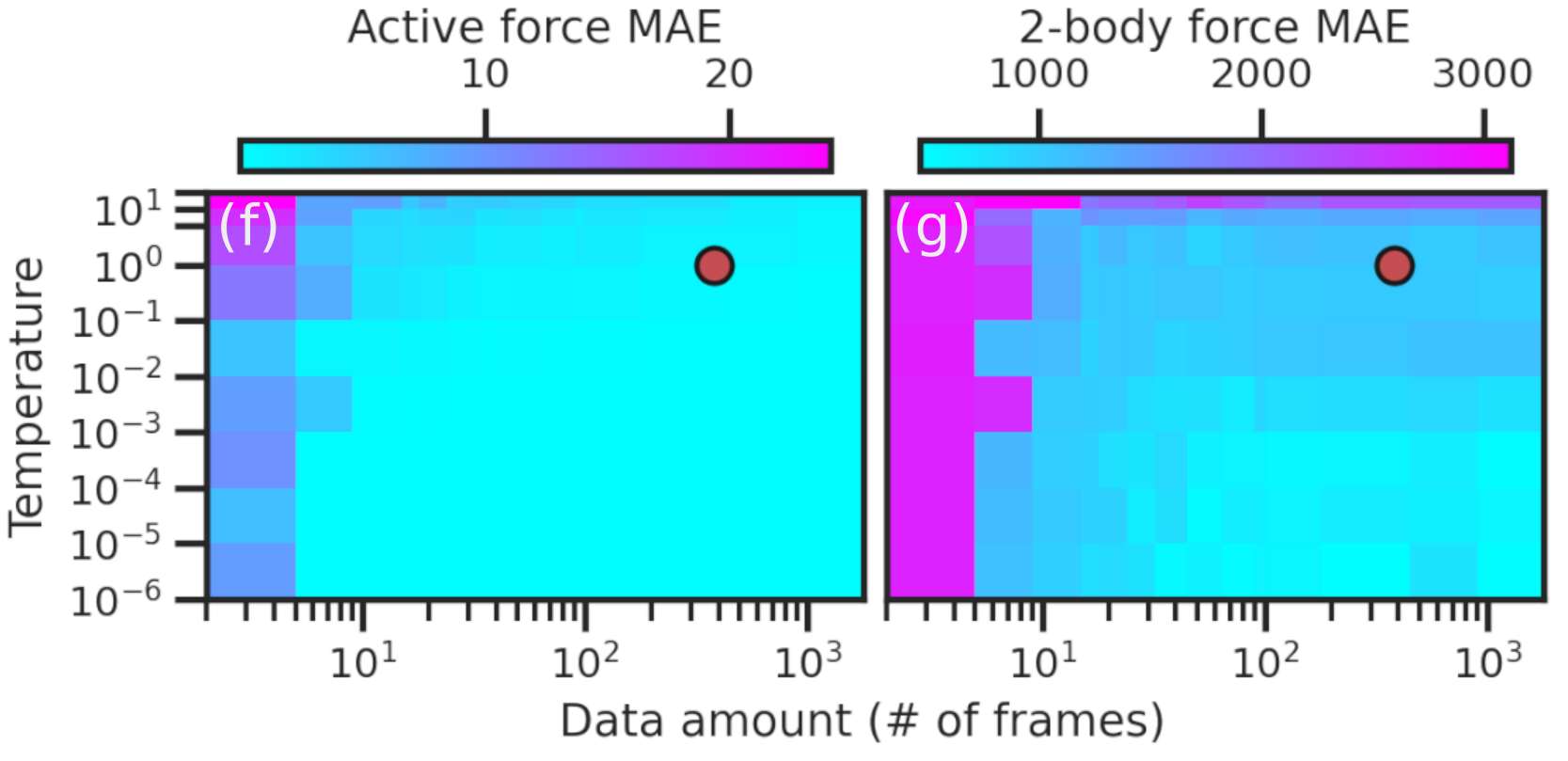}
    \caption{{\bf Mean Absolute Error} for the predicted active and 2-body forces as a function of the temperature of the simulation and the amount of data used for training the network (measured as the number of frames, each containing $2\,500$ particles). The red dot shows the values of these parameters used for figure \ref{fig:gnn_MD}.}
    \label{fig:gnn_MD_err}
\end{figure}

\MBGc{
The results presented in this section can be split into three groups: (i) the application of ActiveNet to simulations of ABPs with different interaction potentials (figure \ref{fig:gnn_MD}, panels a-d); (ii) the validation of ActiveNet with systems in the underdamped regime or presenting torques (figure \ref{fig:gnn_MD}, panels e-f); and (iii) testing the sensibility of the method against thermal noise and data scarcity (figure \ref{fig:gnn_MD_err}).\\
\\
(i) In the overdamped cases, for each simulation ActiveNet is trained using $380$ snapshots of $2\,500$ particles. ActiveNet adopts as input the positions and orientations of all particles in each frame, and learns to predict the correct velocities. In this process, the edge function learns the \MRGb{two-body} force between \textit{any} pair of particles (up to a linear transformation, \MRGb{see appendix section \ref{sec:methods_gnn}}), while the node function learns the active force acting on each particle.
Panels (a), (b), and (c) present the two-body forces learned by ActiveNet, compared with the force used in the simulation (dashed lines). Panel (a) presents a dilute suspension (gas phase) of particles interacting via a repulsive force characterised by two length scales derived from a shoulder potential. Panel (b)
shows a dense suspension of particles that interact via a purely repulsive WCA interaction with high activity. Particles undergo motility-induced phase separation (MIPS), and ActiveNet learns a repulsive force even though the system phase separates due to particles' activity. Panel (c) presents a dilute suspension of particles interacting via an attractive Lennard-Jones interaction with low activity. Particles form ``dynamic'' clusters that jiggle and drift, where very few particles explore different local structures, making it harder for ActiveNet to learn the two-body forces. In this case, ActiveNet slightly underestimates the two-body force; this is likely due to the fact that, at low activities, the two-body forces change within a shorter time scale than in the rest of the cases, leading to a reduction in the correlation between the numerical (average) velocity and the instantaneous force we aim to learn.
It should be noted that the two-body forces learned by ActiveNet dramatically deviate from the true forces present in the simulations for distances smaller than the ones shown in these three panels (a-c). This occurs because at very short distances the repulsion between particles is larger than the active forces acting on them, leading to a lack of data at those distances---in the simulations there are no pairs of particles closer than a certain threshold, controlled by the short-range repulsion and the intensity of the active force. Due to the lack of data, ActiveNet cannot learn the two-body force at distances shorter than the threshold.
In these three systems,  where particles interacted through these two-body forces while  being self-propelled by active forces, ActiveNet was also able to correctly learn the active forces, which are presented in panel (d) for the three systems together.\\
\\
(ii) In the case of underdamped Langevin dynamics, ActiveNet is trained using 94 snapshots of 3\,600 particles each and adopts as input the positions, orientations and velocities of all particles in each frame, and learns to predict the correct accelerations. In this process, the edge function learns the two-body force between \textit{any} pair of particles (up to a linear transformation), as in the Brownian case. However, the node function not only learns the active force acting on each particle but also the drag force, \MRGb{appendix section \ref{sec:methods_gnn} explains how to separate both contributions}. Panel (e) studies a dilute suspension of active particles in the underdamped regime (Langevin dynamics) interacting via a purely repulsive WCA potential with low activity, as shown in the panel, ActiveNet can learn the drag force showing how this method could be used to learn forces that depend on velocities.\\
\\
In the case of chiral ABPs, ActiveNet is trained using 400 snapshots of 3\,600 particles each, it takes as input the positions and orientations of the particles and is trained to predict the linear and angular velocities. In this process, the node function correctly encodes the active force and torque acting on the particles, and the edge function again correctly captures the pairwise repulsive interaction (WCA). The predicted torque is shown in panel (f), while the predicted active and two-body forces can be seen in figure \ref{fig:chiral_learning} of the Appendix section \ref{sec:methods_gnn}.\\
\\
In view of the results for these two last cases, we conclude that the applicability of ActiveNet is thus not restricted to the ABP model and can be applied satisfactorily to underdamped systems in which inertia cannot be disregarded and systems in which there exist additional orientational dynamics.\\
}

\MRGb{(iii) Finally, we study the sensibility of ActiveNet against noisy and scarce data (by increasing the temperature of the simulation and by feeding ActiveNet with subsets of data of decreasing size, respectively). We quantify the error in the forces learned by ActiveNet through the computation of the Mean Absolute Error between the predicted and the ground-truth active and 2-body forces (figure \ref{fig:gnn_MD_err}). Additional details can be found in sections \ref{sec:methods_md} and \ref{sec:mae} (figure \ref{fig:pred_err}). In summary, as one could have expected, higher temperatures and smaller amounts of training data lead to less accurate predictions by ActiveNet. In the cases studied in this work, 10 frames (containing 2\,500 particles each) are enough to learn the forces in a broad range of temperatures.}

\subsection{ActiveNet facilitates the estimation of one-body stochastic forces }

\MRGc{
We have shown how ActiveNet can learn the determinisic forces acting on active brownian particles. However, we may be also interested in discovering the stochastic forces that, together with the deterministic forces, control the dynamics of the system. If there were no deterministic forces, one could always extract the amount of translational or rotational noise in the dynamics of the particles calculating the mean square displacement of such quantities. However, in real cases where activity and interactions are present, the stochastic forces are masked by the interactions and active forces. \\
Fortunately, we can use the deterministic forces learned by ActiveNet to subtract the deterministic contribution to the particles' trajectory. In this way, we obtain virtual trajectories whose dynamics is solely controlled by the stochastic terms. Calculating the Mean Square Displacement on these virtual trajectories we can estimate the stochastic forces present in the dynamics of the particles.\\
Since these virtual particles  diffuse in $x,y,\theta$, subtracting the deterministic contribution learned by ActiveNet from the velocities  (directly extracted from the trajectories) we have access to the stochastic terms:
\begin{align}
    \frac{1}{N_p}\sum_i^{N_p} |v_{x,i}-v^p_{x,i}| &\sim \sqrt{\langle x^2 \rangle/\Delta t^2} \sim \sqrt{2D_t/\Delta t} \label{eq_rw_x}\\
    \frac{1}{N_p}\sum_i^{N_p} |v_{y,i}-v^p_{y,i}| &\sim \sqrt{\langle y^2 \rangle/\Delta t^2} \sim \sqrt{2D_t/\Delta t} \label{eq_rw_y}\\
    \frac{1}{N_p}\sum_i^{N_p} |v_{\theta,i}-v^p_{\theta,i}| &\sim \sqrt{\langle \theta^2 \rangle/\Delta t^2} \sim \sqrt{2D_r/\Delta t}
\end{align}
    where $N_p$ is the number of particles, and we have used the MSD of a random walk in one dimension for each of the components. To showcase this approach, we will study the case of ABPs interacting with a repulsive WCA potential and no torques. In this case, $D_r$ can be directly inferred from $\frac{1}{N_p}\sum_i^{N_p} |v_{\theta,i}|$. However, we cannot extract $D_t$ directly from the trajectories, since the diffusivity would be largely impacted by activity and interactions. We need to use \eqref{eq_rw_x} and \eqref{eq_rw_y} for this purpose, but note that this information is already contained in the loss function that we use to train the model, in this case:
\begin{align}
\label{eq:loss-temp-relation}
    \bar{\mathcal{L}} &= \frac{1}{N_p}\sum_i^{N_p} \left(  |v_{x,i}-v^p_{x,i}|  + |v_{y,i}-v^p_{y,i}| \right) \\
    &\approx \sqrt{8D_t/\Delta t}. \nonumber
\end{align}
$\bar{\mathcal{L}}$ is the \textit{per-particle} loss function.
We now compare  the final value of the loss function computed by ActiveNet during training on the simulations of ABPs for different temperatures (the same ones used in figure \ref{fig:gnn_MD_err}). Fig \ref{fig:loss-temp-relation} shows how the loss computed by ActiveNet is remarkably close to $\sqrt{8 D_t/\Delta t }$, allowing us to estimate $D_t$ directly from the value of the loss.\\
At low temperatures the value of the loss deviates from the theoretical line, although this is only noticeable in log scale (see inset of figure \ref{fig:loss-temp-relation}). At these temperatures, the noise from the dynamics is comparable to the uncertainty in the deterministic forces that ActiveNet has learned. When the network is trained with more data (circular markers in figure \ref{fig:loss-temp-relation}), this deviation decreases. At higher temperatures the loss function follows reasonably well equation (\ref{eq:loss-temp-relation}), although the inset in figure \ref{fig:loss-temp-relation} displays a small deviation where the loss is slightly below equation (\ref{eq:loss-temp-relation}). This could be due to modest overfitting. For systems where torques are present a similar approach can be followed, one should subtract the torques learned by ActiveNet from the angular velocities before computing the MSD and estimating $D_\theta$.\\
%
%
%
%
\begin{figure}
    \centering
    \includegraphics[width=0.9\linewidth]{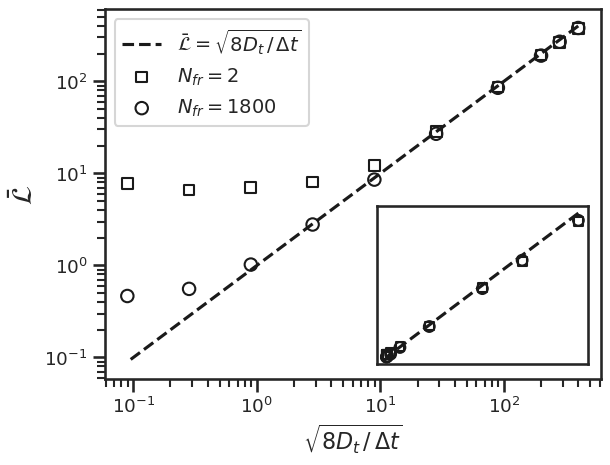}
    \caption{Relation between the value taken by the normalized loss function $\bar{\mathcal{L}}$ and the diffusion constant for the simulations of repulsive ABPs interacting via a repulsive WCA potential. The dashed line shows the theoretical prediction of this relation (see equation \ref{eq:loss-temp-relation}). The data markers show the values of the loss function after 200 epochs of training for two different cases. The empty squares correspond to the loss function of the network trained only with $N_{fr}=2$ frames of a simulation in which $F_p=30$ and $D_r=0.25$. The empty circles correspond to the loss function of the network trained on the same simulation but with $N_{fr}=1800$ frames. The inset shows the same plot with linear scale in both axes.}
    \label{fig:loss-temp-relation}
\end{figure}
}

\subsection{ActiveNet learns active and two-body forces in experiments of electrophoretic Janus particles}
\label{sec:gnn_exp}

Without ActiveNet, extracting the expressions for the active and two-body forces from the collective dynamics of the particles would be extremely hard. ActiveNet allows us to tackle this problem from a completely different point of view: from particles' positions and orientations, ActiveNet learns the active and two-body forces that best predict  particles' velocities. 
\begin{figure}[ht!]
    \centering    
    \includegraphics[width=\linewidth]{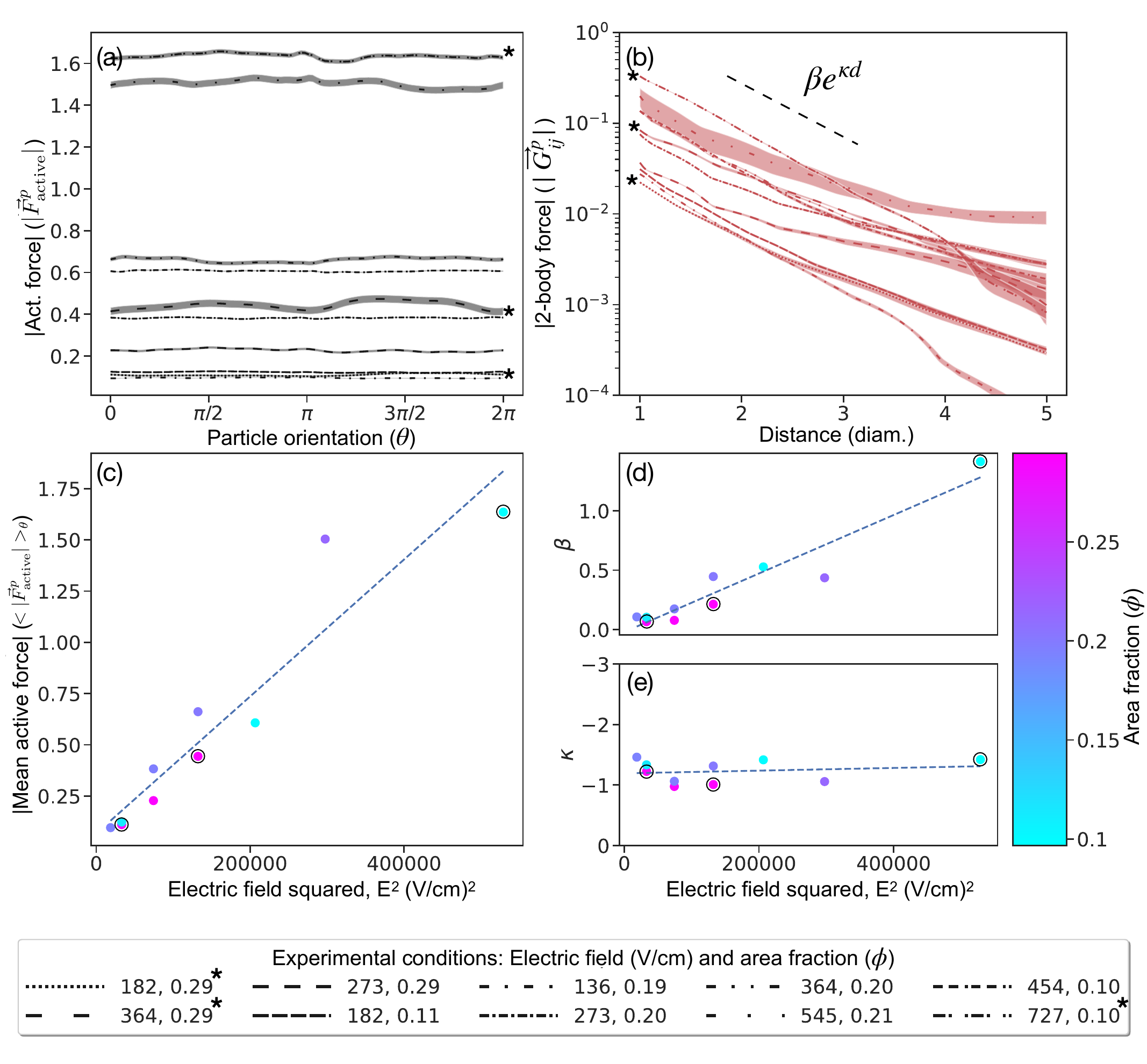}
    \caption{ {\bf ActiveNet can learn the active and inter-particle forces in an experiment of electrophoretic Janus particles} where activity is controlled by the magnitude of the electric field ($E$). \CV{In panels (a) and (b), the ten different series of data (lines with different styles) represent ten different systems, each characterised by a pair of values of the electric field (E) and the area fraction ($\phi$), as reported in the legend, below the four panels. Note that three values of $(E,\phi)$ correspond to the snapshots shown in figure. \ref{fig:exp_snapshots},} \CV{we have marked them with black asterisks in panels (a) and (b) and black circumferences in panels (c), (d) and (e). In panels (a) and (b) we train ten instances of ActiveNet for each dataset with different initial seeds, lines correspond to the average result and the shadows around the lines mark the estimated uncertainty (error bar) by means of the standard deviation.} Panel (a) displays the magnitude of the active force versus the particle's orientation. All lines show horizontal behavior indicating the absence of a preferred direction in the system. Panel (b) displays the modulus of the inter-particle force versus the distance between two particles. All forces are repulsive (red lines) and they approximately follow exponential decays, $\beta e^{\kappa d}$. The black dashed line does not correspond to any particular fit and is included for visualization. Panel (c) reports the average value of the active force in panel (a) versus $E^2$. The dashed line is a linear fit to the points and indicates that the active force is proportional to $E^{2}$. Panels (d) and (e) show $\beta$ and $\kappa$  ($\beta e^{\kappa d}$, $d \in [1,3]$) for all the lines in panel (b) versus $E^2$ (individual fits are not plotted in panel (b)). From panel (d), $\beta$ scales as $E^{2}$, the dashed blue line represents a fit to the data. Panel (e) shows that, $\kappa$ does not depend on $E$ or $\phi$. Its value fluctuates between $-1$ and $-2$ leading to a length scale of the order of one particle diameter. The color of the points in panels (b), (d) and (e) correpond to the area fraction ($\phi$, see colorbar).}
    \label{fig:gnn_exp}
\end{figure}
For simplicity, we build ActiveNet assuming that the active force depends on the orientation of the particle and the two-body force depends on the distance between pairs of particles (see appendix section \ref{sec:methods_edge_node_functions} for more details). \MRG{Note that to use ActiveNet we do not need to assume if the forces are conservative or not.} We train ActiveNet with the data extracted from ten different experiments, performed at different values of both electric field and area fraction, as reported in figure. \ref{fig:gnn_exp} (the values are explicitly indicated in the legend). Each time, ActiveNet is randomly initialized before the training procedure. Depending on the amount of data gathered in each experiment, we use $50-100$ snapshots containing approximately $1\,000-6\,000$ particles each. 

Figure \ref{fig:gnn_exp} (a) presents the modulus of the active force as a function of particles' orientation  ($\theta$). Approximately horizontal lines indicate that ActiveNet is learning an active force with a constant modulus and no preferred direction. \CV{In this panel, (as well as in panel (b), we indicate by means of shadows around the lines the estimated uncertainties (error bars) for the predictions. Each data point corresponds to the average of an ensemble of ten ActiveNet models, trained with the same data and, as mentioned in the introduction, with different initial seeds, the error bar is the standard deviation of the predictions. We notice that the error bars are small compared to the absolute values of the predicted values and support our observation of no preferred direction}. We calculate the average value of each line and plot it in panel (c) as a function of the square of the electric field amplitude. Points are scattered along a straight line, leading to the expected \cite{yan2016reconfiguring} relation $\vec{F}^p \sim C E^2$, see appendix section \ref{sec:methods_gnn}. 
Moreover, points seem to follow different straight lines depending on their area fraction, which leads to a relation $\vec{F}^p \sim C(\phi) E^2$. \CV{The different straight lines may be due to the different sample cells used for each area fraction. Although these cells are very similar, small differences in the spacing between the confining walls could lead to systematic errors in the observed prefactor ($C(\phi)$). Further work is needed to resolve this. It is, however, remarkable that ActiveNet seems sensitive enough to detect this effect.}

Figure \ref{fig:gnn_exp} (b) shows the two-body force \CV{, with error bar marked by a shadow,} that ActiveNet learns as a function of the distance between two particles. The straight lines in the semi-log plot indicate an exponential decay. Low area fraction experiments ($\phi \sim 0.1$) lead to very clear and almost parallel straight lines, whereas experiments at higher area fractions present more variability. The experiments that combine high electric field and high area fraction present an interrupted MIPS phase (see figure. \ref{fig:exp_snapshots}). Thus, it is reasonable to assume that it will be harder for ActiveNet to learn the two-body interactions in those cases: here, particles are closer together and many-body interactions might play a more pronounced role. We fit each line between $d=1$ and $d=3$ to an expression of the form $\beta e^{\kappa d}$ (fits not shown in the plot). Panels (d) and (e) show the best fit for $\beta$ and $\kappa$, for each experiment. The prefactor of the exponential decay ($\beta$) scales also as $E^2$, consistent with the a screened electrostatic interaction \cite{russel1991colloidal,loeb1961electrical}, since the polarization in each colloid is proportional to $E$. On the other hand, $\kappa$ does not seem to depend on $E$ or on area fraction and has an approximate value of $-1.2$ diameter$^{-1}$, corresponding to a characteristic length scale for this interaction of approximately one particle diameter.

\section{Discussion and future work}
\label{sec:discussion}

Our work demonstrates the potential that applying deep learning methods with inductive biases (\textit{a priori} assumptions) has when studying  collective dynamics of suspensions of active colloidal particles. The most important \textit{a priori} assumptions we considered were that 1) all particles followed the same local rules (edge and node functions) and 2) many-body interactions could be neglected. With these assumptions, ActiveNet was not only able to predict the system's dynamics, but could be directly used to uncover forces and torques acting on active particles. We validated our approach by means of numerical simulations in the under and overdamped regime, where we had complete control over the active and conservative forces. We were able to extract the correct expressions for the forces in cases with different levels of activity and different two-body interactions, including forces that depend on velocities -- such as the drag force --, which opens up the possibility of using ActiveNet to learn more complex hydrodynamic forces in the future. For the case of chiral ABPs, we are also able to extract the active force, torque and 2-body interactions. 
We presented a simple example where the active torque was just a constant term as a proof of concept, i.e., it does not depend on the position or orientation of other particles. In other cases, 
ActiveNet could be also used to extract the torque experienced by the particles, which could depend on the particles' position and orientation or the coordinates of nearby particles \cite{dauchot_torque_2019,bricard2015,weber2013}.Moreover, we also demonstrate a procedure by which using ActiveNet's predictions we can estimate stochastic forces present in ABP suspensions. Thus obtaining a good estimation of the full equation of motion of this system.

\MBGd{Hydrodynamic forces are typically many-body, long-ranged, and they may depend on the position, orientation and velocity of several particles. Generalizing ActiveNet for such situations will be an interesting challenge for the future. Although ActiveNet is currently restricted to two-body interactions, it can in principle learn effective two-body hydrodynamic interactions between particles, possibly depending also on the velocities of the particles. For dilute systems, these two-body interactions may give a reasonable good approximation to the actual hydrodynamic interactions \cite{durlofsky_dynamic_1987}.
When many-body hydrodynamics needs to be taken into account, a tentative approach would be to include a $M\times M$ \textit{hyperedge} function \cite{feng_hypergraph_2019} that takes as input the particles' velocities and outputs (upon learning) the M-body forces. This \textit{hyperedge} function could be the analogue to the \textit{resistance matrix} \cite{durlofsky_dynamic_1987} of Stokesian dynamics \cite{brady_stokesian_1988}.} In this case, we believe that symbolic regression will be necessary to understand the analytic expression for the inter-particle interactions \cite{Ouyang2019,Ouyang2018}.

In the case of experiments with electrophoretic Janus particles, ActiveNet found an active force proportional to $E^2$, in agreement with previous studies \cite{nishiguchi2018flagellar}. Furthermore, ActiveNet was sensitive enough to detect subtle differences showing a dependency on area fraction.
Finally, ActiveNet revealed that particles in these experiments were interacting via a pure repulsion, decaying as an exponential with a length scale ($\kappa^{-1}$) that did not depend neither on area fraction nor electric field, and a prefactor that scaled as $E^2$. This force is  consistent with a screened electrostatic interaction between the colloids. 
Contrary to the case of numerical simulations, the two-body force learned from  experimental data started deviating from the described behavior when interrupted MIPS phases developed, suggesting that many-body interactions might start playing an important role in such cases.

 
We believe that ActiveNet can be directly applied to other suspensions of active particles. For example, it would be interesting to use this approach to distinguish between  forces present in systems of healthy or pathological living cells (similarly to \cite{bruckner2021learning}), which may ultimately lead to a diagnostic tool.
\CV{ActiveNet could also be applied to passive systems. In particular, since ActiveNet  uses particles' dynamics and not  structural information, it could be especially useful when a system of passive colloids is out-of-equilibrium (due to a particular initial condition or to external driving).} 
Additionally, we plan  to test if the training of these GNNs can be improved using a dynamical change of its loss function landscape \cite{ruiz2021tilting},  considering a cutoff ($\Gamma(\tilde{t})$) that changes during training. In cases where the dynamics of the system changes during the observation time, it will be interesting to understand how the architecture of ActiveNet affects the capability of forgetting or transferring previous knowledge to new conditions \cite{mccloskey1989catastrophic,ruiz2022model}.


Finally, the choice between ActiveNet and other available methodologies ultimately hinges on the level of detail required for the sought model and the data accessible to the researcher. If particle trajectories are obtainable, our model can offer a reliable description of the microscopic forces at play in the particle dynamics. In contrast, if the researcher only has access to coarse-grained fields, such as average velocities and densities, our method cannot be applied. In such a case, one can consider alternative approaches that derive appropriate partial differential equations~\cite{brunton2016discovering,rudy2017data,both2021deepmod}. 

To conclude, our work opens up new avenues for understanding systems of active particles. Our approach leads to a ready-to-use tool for experimentalists to learn the forces present in their systems \footnote{Readily usable code with working examples will be available at GitHub at the time of publication.}. The extracted forces will shine a light on the physics underlying experimental systems, which in turn could lead to novel numerical and analytical models, undoubtedly leading to new predictions in the field of active matter. 

\MBGd{
\section*{Data Availability Statement}
The codes used for this work are available in a public repository  \cite{gitlab}.
}

\begin{acknowledgments}
Shortly after submitting this manuscript, the authors became aware of a very interesting and recent work related to ours, also based on graph-networks algorithms, to learn the pairwise interaction and model dynamics at particle level \cite{fink-gnn-2022}. Differently from the cited work, we develop a graph-network scheme for learning not only the pairwise interactions but also one-body terms, such as the active or drag forces of self-propelled colloids and the stochastic force. We  stress again the fact that the novelty of the method here proposed is the ability to decouple one-body and two-body (pairwise) contributions.
M.R.-G. thanks Farshid Jafarpour for enriching discussions about stochastic processes. M. R.-G. also thanks Javier Rodr\'iguez, Leticia Valencia and Jos\'e Luis Jorcano for insightful discussions about living active matter. \MBGc{C.M.B.G., C.V. and M.R.-G. thank Jos\'e Mart\'in-Roca for enriching and fruitful discussions about the GNN structure and support with calculations.} M.R.-G. acknowledges support from the Ram\'on y Cajal program RYC2021-032055-I and from the CONEX-Plus program funded by Universidad Carlos III de Madrid and the European Union's Horizon 2020 research and innovation program under the Marie Sk{\l}odowska-Curie grant agreement No.~801538. L.M.G. acknowledges funding from the European Union’s Horizon 2020 research and innovation program under the grant agreement Nº 951786 (\mbox{NOMAD} CoE).  C.V. acknowledges fundings    IHRC22/00002 and 
PID2022-140407NB-C21 from MINECO. \CV{C.M.G.B. thanks HPC-Europa3 for financial support}.\\
\end{acknowledgments}


\appendix

\section{ActiveNet}
\subsection{Inductive biases for the edge and node functions}
\label{sec:methods_edge_node_functions}

The main assumption when we use ActiveNet is that all the particles follow the same local rules (the edge and node functions) and that we neglect many-body interactions. In this work the edge and node functions are two fully connected deep neural networks with two hidden layers of $300$ units \CV{(as shown in the bottom panels of figure 2)}. In addition to this, we add some extra inductive biases. In the case of experiments, we know  there are imperfections in the substrate where few particles might get stuck. 
Thus, directly providing particles' positions  $(x_i,y_i)$  to the edge and node functions would lead ActiveNet to learn this spurious (although real) information. When dealing with the edge function, we directly provide the distance between two particles, instead of feeding ActiveNet with the coordinates and letting it learn $d=\sqrt{(x_1-x_2)^2+(y_1-y_2)^2}$. Next, we multiply by the unitary vector pointing in the direction connecting the two particles. We also tried providing the distance and the unitary vector as input so that ActiveNet would learn the correct direction of the force: this led to equivalent results. 

For the same reason we do not give the position of the particles to the node function, considering as the only input the particle's orientation $(\cos{\theta}, \sin{\theta})$ and the aggregated output of the edge function applied to the pairs ij. \CV{In the underdamped dynamics we also input the velocity ($\vec{v}_i$) to the node function. Note that we could have given also the orientation of the particles or the velocities to the edge function---which learns the two-body interaction---, this could have led to learning an interaction depending on alignment or velocities. However, we preferred to consider at this time} only the distance between the two particles to prioritize the simplest interpretation of the forces learned by ActiveNet.

\subsection{Extracting the active and two-body forces \MBGb{and the torques}}
\label{sec:methods_gnn}

Probably, the most important feature of ActiveNet is that it allows to learn the \CV{forces governing the dynamics of a system of particles,} together with a clear physical interpretation of the results. After training, ActiveNet is able to predict the deterministic dynamics, extracting inter-particle and active forces directly from $\vec{\xi}$ and  $\vec{\psi}$. As explained in the main text,  ActiveNet  disentangles the components of the velocities that can be explained through $\vec{\xi}$ and  $\vec{\psi}$: to obtain the forces we then use Stokes' law. 

For the simulations \CV{of Brownian particles} the expressions are dimensionless and forces and velocities take the same values. We extract the learned active force from ActiveNet as:
\begin{equation}
    \vec{F}\tsub{active}^p(\vec{c}) = \vec{\psi}\left(\vec{c},\vec{0} \right).
    \label{eq_act_force}
\end{equation}
On the other hand, $\vec{\xi}$ learns the two-body interaction between particles. A key point here is that $\vec{\psi}$ takes as input the sum of several outputs of $\vec{\xi}$. Therefore, if the training is successful and ActiveNet learns the dynamics, $\vec{\psi}$ should only perform a linear transformation on $\sum\limits_{|d_{ij}<\Gamma|}\vec{\xi}(\vec{c}_i,\vec{c}_{j})$: in other words,  $\vec{\xi}(\vec{c}_i,\vec{c}_{j})$ learns the force between any two particles, up to a linear transformation. 
In particular, $|\vec{\xi}(\vec{c_i},\vec{c_j})|$ will be the modulus of the inter-particle force multiplied by an arbitrary constant. 
\CV{The fact that the node function cannot make a nonlinear transformation on the edge function, is the theoretical reason that allows us to disentangle both contributions.}

To recover the inter-particle force learned by ActiveNet (plotted in figures \ref{fig:gnn_MD} and \ref{fig:gnn_exp}) we compute the following
\begin{equation} 
\vec{\psi}\left(\vec{c}_i,\vec{\xi}(\vec{c}_i,\vec{c}_{j})  \right) - \vec{F}\tsub{active}^p(\vec{c}_i).
\end{equation}
\CV{In principle, this quantity could depend on the orientation of particle $i$, or on the angle that particles $i$ and $j$ form with the horizontal ($\alpha_{ij}$). To extract only the dependence on the distance between the two particles we integrate out both degrees of freedom,}
\begin{align}
    \label{eq_Gij}
    {G}_{ij}^p &=     \\ \nonumber
    &=\frac{\int_{0}^{2\pi}
    \int_{0}^{2\pi} 
   \left[
\vec{\psi}\left(\vec{c}_i,\vec{\xi}(\vec{c}_i,\vec{c}_{j})  \right) - \vec{F}\tsub{active}^p(\vec{c}_i)
   \right] \cdot \hat{n}_{ij}\,
\textrm{d}\theta \textrm{d}\alpha_{ij}}{2\pi^2},
\end{align}
where $\hat{n}_{ij}$ is the unitary vector connecting particles $i$ and $j$.
In the case of experiments, \CV{assuming Brownian dynamics}, we obtain forces using Stokes' law. We multiply the learned velocities (which are of the order of $1$ particle's diameter s$^{-1}$) by $6\pi \eta r \sim 1.7 \ 10^{-12}$ N, considering the  water viscosity $\eta \sim 10^{-3}$  Pa s and the particle radius $r\sim 4.3 \mu$m.

\CV{In the case of particles undergoing Langevin dynamics (the underdamped case), we study the trajectories generated by the dimensionless equations described in section \ref{sec:methods_md}. We consider an edge function analogous to the overdamped case. However, the node function takes now also the velocity of the particle as another input and we train ActiveNet to predict the acceleration of particle i:}
\begin{equation}
    \vec{a}_i^p \equiv \vec{\psi} \left(\theta_i,\vec{v}_i,\sum\limits_{d_{ij}<\Gamma} \vec{\xi}(\vec{c}_i,\vec{c}_{j}) \right),
\end{equation}
\CV{where $\theta_i$ and $\vec{v}_i$ are the orientation and velocity of particle $i$. Once ActiveNet has been trained, and it can correctly predict particles' accelerations, we take advantage of the isotropicity of the system to compute the active, drag and two-body forces. We integrate out the velocity of the particle to get the active force:}
\begin{equation}
    \vec{F}\tsub{active}^p (\theta) =
    \frac{\int_{v\tsub{min}}^{v\tsub{max}} 
    \int_{0}^{2\pi} 
    \vec{\psi}\left(\theta,v,\Omega_{v},\vec{0} \right) v \textrm{d}v \textrm{d}\Omega_{v}}
    {\pi (v\tsub{max}-v\tsub{min})^2},
\end{equation}
\CV{where $v$ and $\Omega_{v}$ represent the modulus and orientation of particle $i$'s velocity. Similarly, we compute the drag force integrating out the internal orientation of the particle:}
\begin{equation}
    \vec{F}\tsub{drag}^p (\vec{v}) =
    \frac{\int_{0}^{2\pi} 
    \vec{\psi}\left(\theta,v,\Omega_{v},\vec{0} \right) \textrm{d}\theta}
    {2\pi}.
\end{equation}
\CV{Finally, since we are only interested on the dependence of the two-body force on the distance between particles $i$ and $j$, we integrate out the internal orientation and velocity of particle $i$, and also average over the angle between both particles:
}
\begin{align}
    & \vec{F}\tsub{2-body}^p (d) = \\ 
    & = \frac{
    \iiint_{0}^{2\pi} 
    \int_{v\tsub{min}}^{v\tsub{max}} 
    \vec{\psi}\left(\theta,v,\Omega_{v}, \vec{\xi}(d,\alpha_{ij}) \right) v \textrm{d}v \textrm{d}\omega_{v}\textrm{d}\alpha_{ij}\textrm{d}\theta}
    {4\pi^3 (v_{\max}-v_{\min})^2},\nonumber
\end{align}
\CV{where $\theta$, $v$ and $\Omega_{v}$, represent the internal orientation, speed and orientation of the velocity of particle $i$, whereas $d$ and $\alpha_{ij}$ represent the distance and angle determined by the vector that goes from particle $i$ to $j$}.

\begin{figure}[ht!]
    \includegraphics[width=\columnwidth]{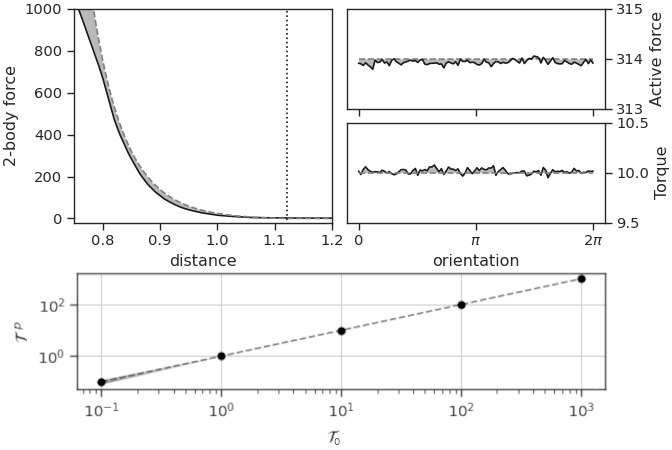}
    \caption{\MBGb{Top left panel: ActiveNet predictions (continuous black line) and ground-truth force (dashed gray) for the 2-body force. Top right panels: torque and active force predictions for the case of $\mathcal{T}_0 = 10$ and ground-truth values (dashed gray). In these panels the shaded area depicts the difference between the predicted and ground-truth values. Bottom panel: predicted ($\mathcal{T}^p$) vs. ground-truth ($\mathcal{T}_0$) torque values (black dots) for the 5 simulations studied and ground-truth reference (dashed gray line). In this panel the shaded area represents the standard deviation of the predicted torques along the given orientations.}}
    \label{fig:chiral_learning}
\end{figure}

\MBGb{In the case of the chiral Active Brownian system, the only modification done to ActiveNet with respect to the previous case of Brownian dynamics was to increase the output dimension of the node function to 3, now $\Vec{\psi}\left(\vec{c}_i,\vec{\xi}(\vec{c}_i,\vec{c}_{j})  \right) = \left(\psi_{v_x}(\dots), \psi_{v_y}(\dots), \psi_\omega(\dots)\right) = \left(v^{\,p}_x, v^{\,p}_y, \omega^p\right)$, where $\omega^p$ is the predicted orientational velocity. We train and test ActiveNet using 400 simulation frames equally spaced throughout the full simulation time for 200 epochs. Similarly, as we did previously for the Brownian dynamics case, the active and 2-body forces are extracted with equations (\ref{eq_act_force}) and (\ref{eq_Gij}) respectively. The torque is extracted with,
\begin{equation}
    \mathcal{T}^{\,p}(\vec{c}\,) = \frac{k_B T}{D_r}\psi_\omega \left( \Vec{c},\, \vec{0} \right)    
\end{equation}
In figure \ref{fig:chiral_learning} we can see the ActiveNet predictions for the the 2-body force (top, left) and the torque and active force  (top right) for the case of $\mathcal{T} = 10$. In figure \ref{fig:chiral_learning} bottom row we can see the predicted vs. ground-truth values for the 5 simulations studied.}

\subsection{Estimating the Mean Absolute Error (MAE)}\label{sec:mae}

\MBG{To estimate the validity range of ActiveNet, we compute the Mean Absolute Error for the predicted active and 2-body forces (equations \ref{ec:mae_acf} and \ref{ec:mae_tbf} ) of particles in a two-dimensional dilute suspension of $2\,500$ active repulsive (WCA) Brownian particles characterised by an active force of  30 (reduced LJ units). }
\begin{align}
 &\text{MAE}(\vec{G}^p_{ij})=\frac{1}{N}\sum_{d}|\vec{G}^p_{ij}(d)-\vec{F}^{\text{theo}}_{\text{WCA}}(d)|\label{ec:mae_tbf}\\
 &\text{MAE}(F^p_x) = \frac{1}{N}\sum_{\theta}|F_x^{p}(\theta)\,-\,|F_p|\cos\theta| \nonumber \\ \nonumber 
&\text{MAE}(F^p_y)=\frac{1}{N}\sum_{\theta}|F_y^{p}(\theta)\,-\,|F_p|\sin\theta|\\ \label{ec:mae_acf}
&\text{MAE}(\vec{F}^p) =\sqrt{\text{MAE}^2(F^p_x)+\text{MAE}^2(F^p_y})
\end{align}

\begin{figure}[ht!]
    \centering    
    \includegraphics[width=\linewidth]{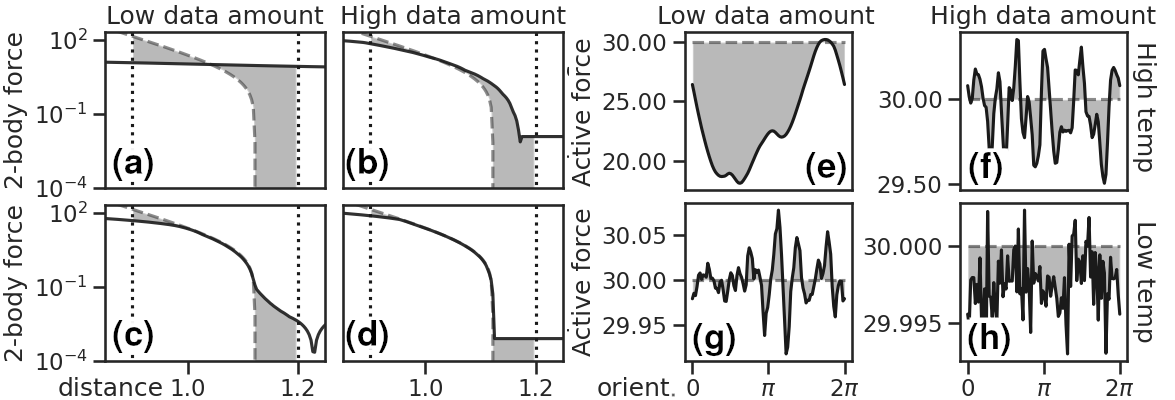}
    \caption{\MBG{Visual representation of the evaluation of the errors. The solid black curves are forces learned by ActiveNet, the dashed curves are the ground-truth forces programmed into the simulation. The shaded region in between them is a measure of ActiveNet's prediction error. The top four panels correspond to the four limit cases for the two-body force: (a) low data -- high temperature, (b) high data -- high temperature, (c) low data -- low temperature and (d) low data -- low temperature. The dotted vertical lines show the range in which the error was computed. The bottom four panels (e,f,g,h) correspond to the same cases for the active force.}
}
    \label{fig:pred_err}
\end{figure}

\MBG{As shown in the main text, \MBGb{figure \ref{fig:gnn_MD_err}} displays the Mean Absolute Error for the predicted active and 2-body forces as a function of the temperature and the amount of data used for training the network (measured as the number of frames). The red dot in these panels shows the value of these parameters used \MBGb{for figure \ref{fig:gnn_MD}}. 
Now, in figure \ref{fig:pred_err} we show the evaluation of errors for the two-body force and the active force. The solid black curves are the forces learned by ActiveNet, the dashed curves are the input forces in the simulation: a WCA potential for the pairwise interactions and an isotropic active force with a modulus of 30 in reduced LJ units.}


\clearpage
\bibliography{scibib}

\end{document}